\documentclass[rapids]{jfm}

\usepackage{graphicx}
\usepackage{epstopdf, epsfig}
\usepackage{dsfont}
\usepackage{psfrag}
\usepackage{amsmath,rotating}
\usepackage{dashrule}
\usepackage{bm}
\usepackage{hyperref} 
\hypersetup{
    colorlinks=true,       
    linkcolor=blue,          
    citecolor=blue,        
    filecolor=blue,      
    urlcolor=black          
}

\usepackage{bigints}
\makeatletter
\@ifpackageloaded{pxfonts}\@tempswafalse\@tempswatrue
\if@tempswa
  \DeclareFontFamily{U}{pxsymbols}{}
  \DeclareFontFamily{U}{pxAMSb}{}
  \DeclareSymbolFont{pxsymbols}{OMS}{pxsy}{m}{n}
  \SetSymbolFont{pxsymbols}{bold}{OMS}{pxsy}{bx}{n}
  \DeclareFontSubstitution{OMS}{pxsy}{m}{n}
  \DeclareSymbolFont{pxAMSb}{U}{pxsyb}{m}{n}
  \SetSymbolFont{pxAMSb}{bold}{U}{pxsyb}{bx}{n}
  \DeclareFontSubstitution{U}{pxsyb}{m}{n}
  \DeclareMathSymbol{\aleph}{\mathord}{pxsymbols}{64}
\fi
\makeatother

\shorttitle{On indirect noise in multicomponent nozzle flows}
\shortauthor{Magri}

\title{On indirect noise in multicomponent nozzle flows}

\author{Luca Magri\aff{1}
  \corresp{\email{lm547@cam.ac.uk}}
 }

\affiliation{\aff{1} University of Cambridge, Engineering Department, Cambridge, UK 
}

\begin{document}

\maketitle
\begin{abstract}
A one-dimensional, unsteady nozzle flow is modelled to identify the sources of indirect noise in multicomponent gases. 
First, from non-equilibrium thermodynamics relations, it is shown that a compositional inhomogeneity advected in an accelerating flow is a source of sound induced by inhomogeneities in the mixture (i) chemical potentials and (ii) specific heat capacities. 
Second, it is shown that the acoustic, entropy and compositional linear perturbations evolve independently from each other and they become coupled through mean-flow gradients and/or at the boundaries. 
Third, the equations are cast in invariant formulation and a mathematical solution is found by asymptotic expansion of path-ordered integrals 
with an infinite radius of convergence. 
Finally, the transfer functions are calculated for a supersonic nozzle with finite spatial extent perturbed by a methane-air compositional inhomogeneity. 
The proposed framework will help identify and quantify the sources of sound in nozzles with relevance, for example, to aeronautical gas turbines. 
\end{abstract}
\section{Introduction}
In order to reduce NO$\textrm{x}$ emissions from aeronautical gas turbines, the objective is to burn in a lean regime. Lean flames burn unsteadily because they are  sensitive to the disturbances in the turbulent environment of the combustion chamber. 
Such unsteady fluctuations in the combustion chamber are the cause of two unwanted phenomena in aero-engines: (i) combustion noise and (ii) thermoacoustic instabilities. 
On the one hand, while methods to mitigate fan and jet noise, which are the primary sources of engine-core noise, have been in place for a decade -- such as ultra-high bypass ratio turbofan engines, acoustic liners and fan blade geometric design -- combustion noise is bound to increase with the implementation of low-emission combustors, high power-density engine cores, and compact burners~\citep[e.g.,][] {DOWLING_MAHMOUDI_PCI2015,Ihme2017}. Combustion noise can cause physiological impairment, such as hearing damage, speech and sleep interference \citep{DOWLING_MAHMOUDI_PCI2015}. 
On the other hand, thermoacoustic instabilities occur when the heat released by the flame is sufficiently in phase with the acoustic waves, which are reflected at the boundaries and generated at the nozzle downstream of the combustion chamber. Thermoacoustic oscillations can cause structural damage and cracking resulting in the reduction of the combustor lifetime by a factor of two or more \citep{DOWLING_MAHMOUDI_PCI2015}.   
Both combustion noise and thermoacoustic instabilities can be caused by direct and indirect mechanisms. 

Direct noise is caused by the acoustics generated by the unsteady heat released by the flame, which is a powerful monopole source of sound.
These acoustics propagate in the combustion chamber through the turbine and are refracted by mean-flow gradients. 
The sound that is transmitted through the downstream-engine component causes noise pollution, 
whereas the sound that is reflected at the nozzle can create a thermoacoustic feedback. 
Indirect noise originates from the acceleration of flow inhomogeneities through the nozzle, or turbine blades, downstream of the combustion chamber. 
The two well-known indirect sources of sound are (i) entropy perturbations (also called ``entropy spots") \citep[e.g.,][]{Cuadra1967,MARBLE_CANDEL_JSV1977,BAKE_RICHTER_MUEHLBAUER_KINGS_ROEHLE_THIELE_NOLL_JSV2009,DURAN_MOREAU_JFM2013} and (ii) vorticity perturbations \citep[e.g.,][]{Howe1977}.
From a thermoacoustic point of view, the acoustics generated at the nozzle by entropy perturbations and travelling back to the combustion chamber can become the key feedback mechanism for a low-frequency combustion instability \citep[e.g.,][]{POLIFKE_PASCHERIET_DOEBBELING_IJAV2001,Goh2013,Motheau2014,Morgans2016}. 
Dissipation and differential convection of entropy perturbations were analysed by Direct Numerical Simulation \citep{MORGANS_GOH_DAHAN_JFM2013} and Large-Eddy Simulation and experiments \citep{Giusti2017}. 
These studies showed that indirect mechanisms, in particular entropy perturbations, need to be modelled for an accurate prediction of indirect combustion noise and thermoacoustic instability in aero-engines. The recent reviews by \citet{DOWLING_MAHMOUDI_PCI2015,Morgans2016} and \citet{Ihme2017} contain an extensive list of references. 

Common to combustion noise and thermoacoustic instability calculations is the necessity to capture in mathematical models all the physical phenomena that contribute to sound generation. 
In the afore-mentioned studies, the perturbations advected with the mean flow were modelled with homogeneous chemical composition. 
In turbulent combustors of aero-engines, however, incomplete mixing and air cooling bring about inhomogeneities in the mixture composition. 
By modelling a multicomponent mixture, \citet{Sinai1980} showed that compositional inhomogeneities, named also ``compositional blobs", generate sound when they are advected through steady low-Mach number flame fronts using the wave equation of \citet{Chiu1974}. 
Recently, it was shown that such compositional blobs contribute to indirect noise in nozzle flows.
This was theoretically shown in compact subsonic nozzles \citep{Ihme2017} and at higher Mach-number regimes with/without shock waves \citep{Magri2016a}. 
By evaluating the nozzle-transfer functions with algebraic expressions \citep{Magri2016a} and numerical integration of the differential equations  \citep{MagriASME2017} of a kerosene mixture, it was shown that compositional noise can be a contributor to indirect noise in lean mixtures and supersonic regimes. 
Compositional noise was experimentally shown by \citet{Rolland2017} as applied to air-helium compositional blobs accelerated through compact-choked nozzles. The compositional and indirect noise they found compared favourably with the noise level predicted by evaluating the transfer functions of \citet{Magri2016a}.   

The objectives of this paper are to 
(i) propose an acoustic model for unsteady multicomponent gases with variable composition and specific heat capacities, for subsonic and supersonic nozzles with finite spatial extent; 
(ii) identify the sources of indirect noise by separating entropic, isentropic and compositional effects;  
(iii) mathematically solve the problem; and 
(iv) numerically quantify the acoustic sources and transfer functions of a supersonic nozzle perturbed by a methane-air blob. 
\section{Mathematical model}\label{sec:mathmodel}
The mixture exiting a combustion chamber of a gas turbine and being accelerated through a nozzle is modelled as 
(i) advection-dominated, where viscosity, heat/species diffusivity and body forces are negligible; 
 (ii) chemically frozen, i.e., the combustion process is completed; 
 (iii) quasi one-dimensional; 
and (iv) isentropic. 
The conservation of mass, momentum, energy and species read, respectively \citep{Chiu1974}
 \begin{align}
 \frac{D{\rho}}{D{t}}+{\rho}\frac{\partial {u}}{\partial {x}}&=0, \\
 {\rho}\frac{D{u}}{D{t}}+\frac{\partial {p}}{\partial {x}}&=0, \\
  T\frac{D{s}}{D{t}}&=- \sum_{i=1}^{N_s}\frac{\mu_i}{W_i}\frac{DY_i}{Dt} + \dot{\mathcal{S}},\label{eq:eneqd}\\
 \frac{DY_i}{D{t}}&=0,   \label{eq:Z}
 \end{align}
which are closed by Gibbs' equation for multicomponent gases
\begin{align}\label{eq:gibbsone}
T{ds} = dh - \frac{dp}{\rho}  - \sum_{i=1}^{N_s}\frac{\mu_i}{W_i}dY_i, 
\end{align}
where 
$\rho$ is the density; 
$u$ is the axial velocity;
$p$ is the pressure; 
$s=\sum_{i=1}^{N_s}s_iY_i$ is the specific entropy;  
$T$ is the temperature; 
$Y_i$ is the mass fraction of the $i$-th species, such that $\sum_{i=1}^{N_s}Y_i=1$, where $N_s$ is the number of species; 
$h=\sum_{i=1}^{N_s}h_iY_i$ is the specific enthalpy; 
${W}_i$ is the molar mass; and 
$\dot{\mathcal{S}}$ is the entropy production by other processes, such as external heat input and enthalpy fluxes, which are assumed negligible in this paper. 
${\mu}_i$ is the chemical potential, which is defined as 
${\mu}_i$ $=$ $W_i\left(\frac{\partial {h}}{\partial Y_i}\right)$ $=$ $W_i\left(\frac{\partial {g}}{\partial Y_i}\right)$, 
where $g=h-Ts$ is the specific Gibbs' energy. 
By using the 1$^{st}$-order homogeneity of Gibbs' energy, which is a thermodynamic potential, it follows that $\mu_i=W_ig_i$. For mixtures of ideal gases, $\mu_i(p,T)=\mu_i^{\circ} +  \mathcal{R}_u T \log(X_ip/p^{\circ})$, where $X_i=(W/W_i)Y_i$ is the mole fraction, $\mathcal{R}_u$ is the universal gas constant and $\circ$ is the reference condition. 
By noting that $\mu_i^{\circ}=g_i^{\circ}=h_i^{\circ}-Ts_i^{\circ}$, \eqref{eq:gibbsone} holds both for the sensible and sensible-plus-chemical entropy, enthalpy and Gibbs' energy, in agreement with \citet{Brear2012}. In this paper, the sensible-plus-chemical quantities will be used. 
The gas is assumed ideal with state equation 
\begin{align}\label{eq:ffssaa}
& p = \rho R T, 
\end{align}
where $R= \mathcal{R}_u\sum_{i=1}^{N_s}Y_i/W_i$ is the mixture specific gas constant. 
The gas is assumed calorically perfect such that $h=c_p(T-T^{\circ})$, 
with $c_p=\sum_{i=1}^{N_s}c_{p,i}Y_i$ being the mixture specific heat capacity at constant pressure, where $c_{p,i}$ is constant. 
Although not necessary, for brevity, the gas composition is parameterized in the mixture fraction space, $Y_i=Y_i(Z)$, hence $dY_i=(dY_i/dZ)dZ$ \citep[e.g.,][]{WILLIAMS_BOOK1985}. 
By considering \eqref{eq:ffssaa}, Gibbs' equation for calorically perfect multicomponent gas becomes  
\begin{align}\label{eq:gibbstwo}
& \frac{ds}{c_p} = \frac{dp}{\gamma p}- \frac{d\rho}{\rho} 
- \left(\aleph+\Psi\right)dZ, 
\end{align}
where $\gamma={c}_p/{c}_v$ is the heat-capacity ratio,
 $c_v$ is the mixture specific heat capacity at constant volume, and  
the relation for ideal gases $R=c_p(\gamma-1)/\gamma$ was used. 
The non-dimensional terms 
\begin{align}
\Psi & =\frac{1}{c_p T}\sum_{i=1}^{N_s}\left(\frac{\mu_i}{W_i}\textcolor{black}{- \Delta{h}^{\circ}_{f,i}}\right)\frac{dY_i}{dZ}, \nonumber\\
\aleph & = \sum_{i=1}^{N_s}\left(\frac{1}{(\gamma-1)}\frac{d\log(\gamma)}{dY_i}+ \textcolor{black}{\frac{T^{\circ}}{T}}\frac{d\log(c_{p})}{dY_i} \right)\frac{dY_i}{dZ}, \label{eq:dunnodd}
\end{align}
are named {\it chemical potential function} and {\it heat-capacity factor}, respectively. If the sensible quantities were used in \eqref{eq:gibbsone}, the term $\Delta{h}^{\circ}_{f,i}$ in \eqref{eq:dunnodd} would not appear.  
\subsection{Linearization}\label{sec:linearization}
A generic  variable is split as $(\cdot)=\bar{(\cdot)}+(\cdot)'$, where $\bar{(\cdot)}$ is the steady mean-flow component, 
and $(\cdot)'\sim O(\epsilon)$ is the unsteady fluctuation with $\epsilon\rightarrow0$. 
By grouping the steady terms, the equations for the mean flow read 
\begin{align}
d(\bar{\rho}\bar{u})=0,\quad d\left(\bar{p} + \bar{\rho}\bar{u}^2\right)=0,\quad d\bar{s}=d\log\left(\bar{p}^{\frac{1}{\bar\gamma}}/\bar{\rho}\right)=0,\quad d{\bar{Z}} =0.  
\end{align}
The third relation shows that the mean flow has constant and uniform entropy, i.e., it is homentropic.  
The mean-flow specific heat-capacity ratio, $\bar{\gamma}$, and specific heat capacity, $\bar{c}_p$, are constant because they depend only on $\bar{Z}$, i.e., $d\bar{\gamma}=0$ and $d\bar{c}_p=0$. 
The variables are non-dimensionalized as ${\eta}={x}/{L}$, where ${L}$ is the nozzle axial length; 
$\tau={t}{f}$ where ${f}$ is the frequency at which the advected perturbations enter the nozzle;
$\bar{M}=\bar{u}/\bar{c}$ is the mean-flow Mach number;
and $\tilde{u}=\bar{u}/\bar{c}_{ref}$, where $\bar{c}_{ref}$ is a reference speed of sound. 
On linearization of \eqref{eq:gibbstwo} and taking the material derivative, Gibbs' equation reads 
\begin{align}
& \frac{\bar{D}}{D\tau}\left(\frac{p'}{\bar{\gamma}\bar{p}} - \frac{\rho'}{\bar{\rho}} - \frac{s'}{\bar{c}_p}\right) 
- \left(\bar{\Psi} + \bar{\aleph}\right)\frac{\bar{D}Z'}{D\tau} 
- \frac{\bar{D}\bar{\Phi}}{D\tau}Z' = 0,  \label{eq:gibbsthree}
\end{align}
where 
\begin{align}\label{eq:phistr}
\bar{\Phi}=\frac{d\log(\bar{\gamma})}{dZ}\log(\bar{p}^{\frac{1}{\bar{\gamma}}})
\end{align}
is the $\gamma$-source of noise, named by \citet{Strahle1976}.  
The linearized material derivative is defined as $\bar{D}(\cdot)/D\tau=He\partial(\cdot)/\partial\tau + \tilde{u}\partial(\cdot)/\partial \eta$, where 
 $He={f}{L}/\bar{c}_{ref}$ is the Helmholtz number, which is the ratio between the advected-perturbation and acoustic wavelenghts. 
%
%
Equation \eqref{eq:gibbsthree} can be integrated 
from an unperturbed condition along the characteristic line $He\tau=\int^\eta d\tilde{\eta}/\tilde{u}(\tilde{\eta})$, to yield the density fluctuation 
 \begin{align}
& \frac{\rho'}{\bar{\rho}} = \frac{p'}{\bar{\gamma}\bar{p}}  - \frac{s'}{\bar{c}_p}- \left(\bar{\Psi} + \bar{\aleph} + \bar{\Phi}\right)Z', \label{eq:eqgibbsint}
\end{align}
where $\bar{\Psi}$ and $\bar{\aleph}$ are evaluated where the compositional inhomogeneity is generated,
while $\bar{\Phi}$ is a spatial function. 
In \eqref{eq:eqgibbsint}, the fact that $Z'=Z'\left\{He\tau - \int^\eta \frac{d\tilde{\eta}}{\tilde{u}(\tilde{\eta})} \right\}$ and $s'=s'\left\{He\tau - \int^\eta \frac{d\tilde{\eta}}{\tilde{u}(\tilde{\eta})} \right\}$ are Riemann invariants was exploited, as shown later in \eqref{eq:eq13}-\eqref{eq:eq14}. 
Equation \eqref{eq:eqgibbsint} physically signifies that the excess density, $\rho'/\bar{\rho}-p'/(\bar{\gamma}\bar{p})$ \citep{MORFEY_JSV1973}, varies because of 
(i) entropic perturbations, $s'/\bar{c}_p$ and 
(ii) compositional perturbations, 
whose strength is related to the chemical potential function, $\bar{\Psi}$; 
(iii) and isentropic perturbations due to gas heat capacity variations through $\bar{\aleph}$ and $\bar{\Phi}$.  When species are generated/added upstream of the inlet, they also bring in an entropic contribution through the chemical potential, which can be calculated from the energy equation \eqref{eq:eneqd}. 
Finally, by grouping the terms of order $\sim O(\epsilon)$, the  linearized multicomponent gas equations read  
\begin{align}
& \textcolor{black}{\frac{\bar{D}}{D\tau}\left(\frac{p'}{\bar{\gamma}\bar{p}}\right) + \tilde{u}\frac{\partial}{\partial \eta}\left(\frac{u'}{\bar{u}}\right) - \frac{\bar{D}\bar{\Phi}}{D\tau}Z'=0}, \label{eq:eq10}\\
&\textcolor{black}{\frac{\bar{D} }{D\tau}\left(\frac{u'}{\bar{u}}\right) +  \frac{\tilde{u}}{\bar{M}^2}\frac{\partial}{\partial \eta}\left(\frac{p'}{\bar{\gamma}\bar{p}}\right)}+\Bigg[2\frac{u'}{\bar{u}} + \textcolor{black}{(1- \bar{\gamma})\frac{p'}{\bar{\gamma}\bar{p}}} -\frac{s'}{\bar{c}_p} - \left(\bar{\Psi}+ \bar{\aleph} + \bar{\Phi}\right)Z' 
\Bigg]\frac{d \tilde{u}}{d\eta} =0,\label{eq:eq12} \\   
& \frac{\bar{D}}{D\tau}\left(\frac{s'}{\bar{c}_p}\right) = 0, \label{eq:eq13}\\
& \frac{\bar{D}Z'}{D\tau}= 0.   \label{eq:eq14}
\end{align}
Equation \eqref{eq:eq12} shows the sources of indirect noise. 
The unsteady interaction between the specific entropy perturbation, $s'$, the compositional perturbation $Z'$ and the mean-flow gradient, $d\tilde{u}/d\eta$, is the source of indirect noise. 
Physically, not only do density variations create noise through entropy mechanisms, but also differences in species generate noise through the  chemical potential \citep{Magri2016a} and heat-capacity variation. 
Such a sound generation is caused by the tendency of the compositional blob to contract/expand with a different rate than the surrounding fluid. 
Equation \eqref{eq:eq13} shows that the linearized flow has constant entropy along a pathline, i.e., it is isentropic but not necessarily homentropic. 
\subsection{Riemann invariants and flow modes}
\begin{figure}
  \begin{center}
\includegraphics[width = 0.8\columnwidth,draft=false]{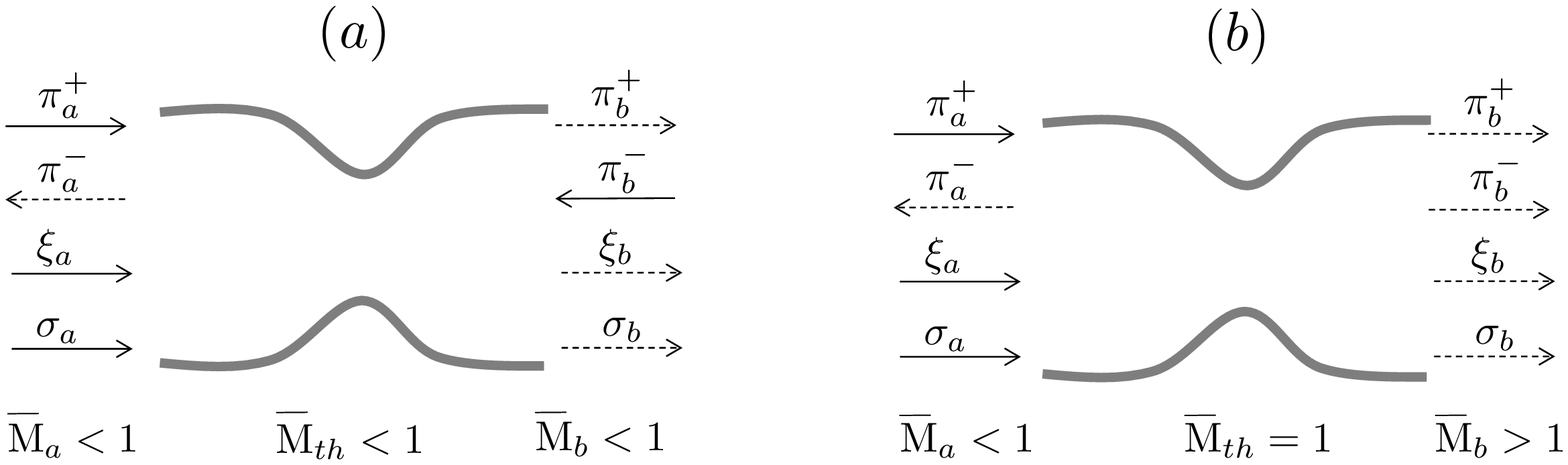}
   \caption{\label{FIG_COMPACT_NOZZLE}Pictorial decomposition of acoustic wave, $\pi$, specific entropy inhomogeneity, $\sigma$, and compositional blob, $\xi$,  in a (a) subsonic nozzle and (b) supersonic nozzle. $\bar{M}_{th}$ is the Mach number at the throat. }
   \end{center}
\end{figure}

Equations \eqref{eq:eq10}-\eqref{eq:eq14} can be recast in matrix form as $He\partial\mathbf{q}/\partial\tau=-\tilde{u}\mathbf{H}_1\partial\mathbf{q}/\partial \eta + \mathbf{H}_2\mathbf{q}$,  
where $\mathbf{q}=\left[p'/(\gamma\bar{p}),u'/\bar{u},s'/\bar{c}_p,Z'\right]^T$. Matrix $\mathbf{H}_1$ is eigendecomposed as 
\begin{align}\label{eq:eigendec}
\mathbf{H}_1=\mathbf{Q}\bm{S}\mathbf{Q}^{-1} = 
\small{\left[
\begin{array}{cccc}
 {1}& 1& 0 & 0  \\
   \frac{1}{\bar{M}} &-\frac{1}{\bar{M}}  &  0 & 0  \\
 0 & 0 &  1 & 0   \\
   0 & 0 &  0 & 1
\end{array}
\right]
\left[
\begin{array}{cccc}
  \frac{ \bar{M}+1}{\bar{M}} &  0 & 0 & 0 \\
  0 &  \frac{ \bar{M}-1}{\bar{M}}  & 0                    & 0 \\
  0 & 0  & 1 & 0  \\
  0 & 0  &   0 & 1\\
\end{array}
\right]
\left[
\begin{array}{cccc}
 \frac{1}{2}& \frac{\bar{M}}{2}& 0 & 0  \\
   \frac{\bar{1}}{2} &-\frac{\bar{M}}{2}  &  0 & 0  \\
 0 & 0 &  1 & 0   \\
   0 & 0 &  0 & 1
\end{array}
\right]}. 
\end{align}
There are four Riemann invariants: 
(i) upstream and downstream propagating acoustic modes (first two rows of $\mathbf{Q}^{-1}$, $\pi^\pm = \frac{1}{2}\left(\frac{p'}{\bar{\gamma} \bar{p}}\pm \bar{M}\frac{u'}{\bar{u}}\right)$), 
(ii) an advected entropy mode (third row of $\mathbf{Q}^{-1}$,  $\sigma = \frac{s'}{\bar{c}_p}$), 
and (iii) an advected compositional mode (fourth row of $\mathbf{Q}^{-1}$, $\xi = Z'$). By assuming a homogeneous medium, i.e., $\mathbf{H}_2=0$, and a Fourier transform in space, i.e., $\mathbf{q}(\tau,\eta)=\hat{\mathbf{q}}(\tau)\exp(-\mathrm{i}\kappa \eta)$, it can be seen that these modes evolve independently of each other and they become coupled at the boundaries and through mean-flow gradients. 
(The entropic and compositional modes would be coupled through diffusion and reaction effects, which are neglected under the assumptions made.)
The eigendecomposition \eqref{eq:eigendec} justifies a characteristic decomposition of the governing equations, in which four invariants are defined at each side of the nozzle, as shown in Figure~\ref{FIG_COMPACT_NOZZLE}.  
%
%
%
\section{Invariant formulation}
The variables are expressed as functions of the invariants of the compact-nozzle solution ($He=0$). 
When $He=0$, the normalized mass flow rate, $I_{\dot{m}}$, total specific enthalpy, $I_{h_T}$, specific entropy, $I_s$, and mixture fraction, $I_Z$, 
are conserved. 
The total specific enthalpy is the sum of the specific enthalpy and the specific kinetic energy, i.e., $h_T=h+1/2u^2$. 
The specific entropy and compositional invariants, $I_s$ and $I_Z$,  do not depend on mean-flow gradients, therefore they are conserved also in the non-compact nozzle. 
The flow invariants $\bm{\mathcal{I}}=\left[I_{\dot{m}}, I_{h_T}, I_{s}, I_z \right]^T$ are related to the Riemann invariants, $\mathbf{r}=(\pi^+,\pi^-,\sigma,\xi)$ as $\bm{\mathcal{I}} = \mathbf{D}\mathbf{r}$, where  
\begin{align}
&\mathbf{D}=
\left[
\begin{array}{cccc}
 \frac{\bar{M}+1}{\bar{M}} & \frac{\bar{M}-1}{\bar{M}}  & -1 & -\left(\bar{\Psi}+ \bar{\aleph}\right)  \\
\frac{(\bar{\gamma}-1)(\bar{M}+1)}{\beta} & \frac{(\bar{\gamma}-1)(1-\bar{M})}{\beta}  &   \frac{1}{\beta} & \frac{\bar{\Psi}}{\beta}\\
 0 & 0  & 1 & 0 \\
  0 & 0  & 0 & 1 
\end{array}
\right], 
\end{align}
where $\beta=1+\frac{\bar{\gamma}-1}{2}\bar{M}^2$.  
By considering that $d\tilde{u}/d\eta = \tilde{c}\left(1 +\frac{\bar{\gamma}-1}{2}\bar{M}^2\right)^{-1}d\bar{M}/d\eta$, and  
when Fourier-transformed as $\bm{\mathcal{I}}(\tau,\eta)=\hat{\bm{\mathcal{I}}}(\eta)\exp\left(2\pi \mathrm{i}\tau\right)$,  the equations can be cast in matrix from as
\begin{align}\label{eq:systAf}
&2\pi\mathrm{i}He{\hat{\bm{\mathcal{I}}}}= \mathbf{E}(\eta)\frac{d\hat{\bm{\mathcal{I}}}}{d\eta } \textcolor{black}{+\mathbf{F}(\eta)\hat{I}_Z}, 
\end{align}
where 
\begin{align}
\mathbf{E}(\eta) & =
-\tilde{u}\left[
\begin{array}{cccc}
 1 & \frac{\beta}{(\bar{\gamma}-1)\bar{M}^2}  & \frac{-1}{(\bar{\gamma}-1)\bar{M}^2} &  \frac{-\bar{\Psi}}{(\bar{\gamma}-1)\bar{M}^2}  \\
 \frac{\bar{\gamma}-1}{\beta} & 1  &   \frac{\bar{\gamma}-1}{\beta} & \frac{\bar{\gamma}-1}{\beta}\left(\bar{\Psi}+\bar{\aleph}\right)\\
 0 & 0  & 1 & 0 \\
  0 & 0  & 0 & 1 
\end{array}
\right],\\
\mathbf{F} & =  \left[ \tilde{u}\frac{d\bar{\Phi}}{d\eta}+\bar{\Lambda},\;\; \frac{\bar{\gamma}-1}{\beta}\left(\tilde{u}\frac{d\bar{\Phi}}{d\eta} + \bar{M}^2\bar{\Lambda}\right), \;\;0,\;\;0\right]^T, \label{eq:fvect}\\
\bar{\Lambda} & = 
\frac{ {\tilde{u}}\left(\bar{\aleph}+\bar{\Phi}\right)}{\bar{M}^2(\bar{\gamma}-1)}\frac{d\log\left(1+\frac{\bar{\gamma}-1}{2}{\bar{M}^2}\right)}{d{\eta}}. 
\end{align}
This set of equations tends to the single-component gas equations of \citet{DURAN_MOREAU_JFM2013} as $\hat{I}_Z\rightarrow 0$. 
The boundary conditions may be prescribed using the procedure proposed in \S5 of \citet{DURAN_MOREAU_JFM2013}, which is extended to include the additional invariant, $\hat{I}_Z$. 
Physically, the perturbations prescribed at the nozzle's inlet are generated in the combustion chamber due to the unsteady and inhomogeneous combustion process.
\section{Solution}\label{sec:solution}
A solution of the multicomponent acoustic problem \eqref{eq:systAf} is proposed when 
$\mathbf{F}=0$.  
(In \S\ref{sec:acsources}, it is shown that this term is negligible as compared to the other acoustic sources.)
The solution holds for all nozzle locations except where the flow becomes sonic, $\eta=\eta^*$. 
The treatment of the sonic condition is explained in \S\ref{sec:soniccond}. 
First, \eqref{eq:systAf} is recast as 
\begin{align} \label{eq:systAA}
2\pi\mathrm{i}He \mathbf{A}(\eta)\hat{\bm{\mathcal{I}}}= \frac{d \hat{\bm{\mathcal{I}}}}{d\eta}, 
\end{align}
where $\bar{M}\not=1$, and $\mathbf{A}=\mathbf{E}^{-1}$ reads  
\begin{align}\label{eq:systAfA}
&\mathbf{A}(\eta)=\nonumber\\
&-\frac{1}{\tilde{u}}\left[
\small{\small{\begin{array}{cccc}
 \frac{\bar{M}^2}{\bar{M}^2-1} & -\frac{\beta}{(\bar{\gamma}-1)(\bar{M}^2-1)}  & \frac{\bar{\gamma}}{(\bar{\gamma}-1)(\bar{M}^2-1)} &  \frac{\left(\bar{\Psi}+ \bar{\aleph}\right)\bar{\gamma}-\bar{\aleph}}{(\bar{\gamma}-1)(\bar{M}^2-1)}  \\
-\frac{(\bar{\gamma}-1)\bar{M}^2}{(\bar{M}^2-1)\beta} &    \frac{\bar{M}^2}{\bar{M}^2-1}  &  -\frac{1+(\bar{\gamma}-1)\bar{M}^2}{(\bar{M}^2-1)\beta} & -\frac{\left(\bar{\Psi}+ \bar{\aleph}\right)\bar{M}^2\left(\bar{\gamma}-1\right) + \bar{\Psi}}{(\bar{M}^2-1)\beta}\\
 0 & 0  & -1 & 0 \\
  0 & 0  & 0 & -1 
\end{array}}}
\right].
\end{align}

Equation \eqref{eq:systAA} is a set of four linear ordinary differential equations with spatially dependent coefficients. (Note that the system can be reduced to a $2\times 2$ system by using the analytical expressions for the advected entropy spot and compositional blob \textcolor{black}{$I_{s,Z}\left\{He\tau-{\int^{\eta}{\frac{d\tilde{\eta}}{\tilde{u}(\tilde{\eta})}}}\right\}=\hat{I}_{s,Z}\exp\left(-2\pi \textrm{i}He{\int^{\eta}{\frac{d\tilde{\eta}}{\tilde{u}(\tilde{\eta})}}}\right)$.})
If the acoustic commutator of $\mathbf{A}$ is nil, i.e., $\left[\mathbf{A}(\eta_1),\mathbf{A}(\eta_2)\right]=\mathbf{A}(\eta_1)\mathbf{A}(\eta_2) - \mathbf{A}(\eta_2)\mathbf{A}(\eta_1)=0$, for example when $\mathbf{A}$ is a scalar or a constant matrix, the solution of \eqref{eq:systAA} is
$\hat{\bm{\mathcal{I}}}=\exp\left(2\pi\mathrm{i}He\int_{\eta_a}^{\eta}\mathbf{A}(\eta')d\eta'\right)\hat{\bm{\mathcal{I}}}_a$,
where $\exp(\cdot)$ is the matrix exponential. 
However, if the commutator is not zero, for example in the acoustic flow of this paper, the matrix-exponential solution no longer holds, because $\exp\left(\mathbf{A}(\eta_1)\right)\exp\left(\mathbf{A}(\eta_2)\right)\not=\exp\left(\mathbf{A}(\eta_1)+\mathbf{A}(\eta_2)\right)$.
A solution for this case is derived by asymptotic expansion. 
The differential equation \eqref{eq:systAA} is recast in integral form as 
\begin{align}\label{eq:forsolA}
\hat{\bm{\mathcal{I}}}(\eta) = \hat{\bm{\mathcal{I}}}_a+ 2\pi\mathrm{i}He\int^{\eta}_{\eta_a}\mathbf{A}(\eta')\hat{\bm{\mathcal{I}}}(\eta') d\eta', 
\end{align}
which enables an explicit expression for the solution by recursion. 
First, the case $He=0$ is solved, which is a compact nozzle denoted by the subscript $a$. 
From \eqref{eq:systAA}, the solution $\hat{\bm{\mathcal{I}}}_a$ is constant and is equal to its value at the inlet $a$. 
Second, by recognizing the Helmholtz number as the perturbation parameter, the solution is expanded as 
\begin{align}\label{eq:expansion}
& \hat{\bm{\mathcal{I}}} = \hat{\bm{\mathcal{I}}}_{a} + \sum_{n=1}^{\infty}He^n\hat{\bm{\mathcal{I}}}_{n}. 
\end{align}
The asymptotic decomposition \eqref{eq:expansion} is substituted into \eqref{eq:systAA} and, using \eqref{eq:forsolA} by recursion, a solution is derived as follows
\begin{align}\label{eq:recursion_U}
&\small{\hat{\bm{\mathcal{I}}}(\eta) = }\nonumber\\
& \small{\underbrace{\Bigg[\mathds{1} + 2\pi \mathrm{i}He\int^{\eta}_{\eta_a}d\eta^{(1)}\mathbf{A}\left(\eta^{(1)}\right) 
+\ldots+ ( 2\pi \mathrm{i}He)^n\int^{\eta}_{\eta_a}d\eta^{(1)}\ldots\int_{\eta_a}^{\eta^{(n-1)}} d\eta^{(n)}\mathbf{A}\left(\eta^{(1)}\right)\dots\mathbf{A}\left(\eta^{(n)}\right)\Bigg]}_{\textrm{Acoustic propagator}, \;\;\; \mathbf{U}=\mathds{1} + \sum_{n=1}^{\infty}\left(2\pi\mathrm{i}He\right)^n\mathbf{P}_n}\hat{\bm{\mathcal{I}}}_a}, 
\end{align}
where $\eta_a<\eta^{(n)}<\ldots<\eta^{(1)}<\eta$, and $\mathds{1}$ is the identity operator. The integral operators in \eqref{eq:recursion_U} are path-ordered, which means that the operator closer to the nozzle inlet, $\eta=\eta_a$, is always on the right of the operator acting at a farther location. 
Equation \eqref{eq:recursion_U} contains the Neumann series of the acoustic propagator $\mathbf{U}=\mathds{1} + \sum_{n=1}^{\infty}\left(2\pi\mathrm{i}He\right)^n\mathbf{P}_n$, defined as the map such that $\hat{\bm{\mathcal{I}}}(\eta) = \mathbf{U}(\eta)\hat{\bm{\mathcal{I}}}_a$. 
Although asymptotic, solution \eqref{eq:recursion_U} is absolutely convergent in a finite spatial domain and when $\mathbf{A}$ is bounded, which is the case here. 
This can be shown by defining the path-ordering operator 
 $\mathcal{P}\left(\mathbf{P}(\eta_1),\mathbf{P}(\eta_2)\right) = \mathbf{P}(\eta_1)\mathbf{P}(\eta_2)$ if $\eta_1>\eta_2$ and
 $\mathcal{P}\left(\mathbf{P}(\eta_1),\mathbf{P}(\eta_2)\right)=\mathbf{P}(\eta_2)\mathbf{P}(\eta_1)$, if $\eta_2>\eta_1$. 
After some algebra, it can be shown that \citep{Lam98}
\begin{align}
\hat{\bm{\mathcal{I}}} = \mathcal{P}\left(2\pi \mathrm{i}He\int^{\eta}_{\eta_a}\exp\left(\mathbf{A}\left(\eta'\right)\right)d\eta'\right)\hat{\bm{\mathcal{I}}}_a, 
\end{align}
which means that
\begin{align}
\lVert \hat{\bm{\mathcal{I}}} \rVert < \exp\left(\int_{{\eta_a}}^{\eta}d\eta'\lVert\mathbf{A}(\eta')\rVert\right)\lVert\hat{\bm{\mathcal{I}}}_a\rVert<\infty
\end{align}
is absolutely convergent.
Once the acoustic propagator, $\mathbf{U}$, is calculated, 
(i) the solution can be calculated without iterative shooting methods for any boundary conditions; and 
(ii)  the effect of the Helmholtz number can be directly isolated at each order. 
In time-dependent perturbations of quantum systems, the acoustic solution \eqref{eq:recursion_U} has an analogy to the Dyson series, while the integrands have analogies to the Feynman path integrals \citep{Dyson49}. With solution \eqref{eq:recursion_U}, the nozzle transfer functions can be calculated (\S\ref{sec:acsources}). 
\subsection{Sonic conditions}\label{sec:soniccond}
When the flow is choked, the subsonic-flow perturbations calculated at $\bar{M}=1-\epsilon$ provide the input to the supersonic flow at $\bar{M}=1+\epsilon$, where $\epsilon\ll1$, as proposed by  \citet{DURAN_MOREAU_JFM2013} in single-component nozzle flows. In this small range, the local Helmholtz number is negligible and the compact-nozzle solution holds. 
The upstream propagating acoustic wave is given by the reflection at the throat of the incoming perturbations, as $\pi^{-}_{th}=R_{\pi,th}\pi^{+}_{th} + R_{s,th}\sigma + R_{\xi,th}\xi$. 
Taking the limit $\bar{M}\rightarrow1$ of the compact-nozzle transfer functions, derived by setting $He=0$ and $\mathbf{F}=0$ in \eqref{eq:systAf}, provides the reflection coefficients at the choked throat 
\begin{align}
R_{\pi,th} = \frac{3-\bar{\gamma}}{1+\bar{\gamma}}, \quad R_{s,th}= \frac{-1}{1+\bar{\gamma}}, \quad \textcolor{black}{R_{\xi,th}=-\frac{\bar{\Psi}+ \textcolor{black}{\bar{\aleph}}}{1+\bar{\gamma}}}. \label{eq:fihjfrnrfuir}
\end{align}
The first two reflection coefficients were derived by \citet{MARBLE_CANDEL_JSV1977}, and $R_{\xi,th}$ appears in multicomponent gases. 
$R_{\xi,th}$ can be used to assess the effect that the compositional blob has on thermoacoustic instability.  
The choking condition at the throat $M'/\bar{M}=0$  provides the input perturbations to the supersonic part of the nozzle. In terms of Riemann invariants, this condition reads 
\begin{align}
& \frac{1}{2}\sigma+\left(\frac{\bar{\Psi}+ \bar{\aleph}}{2}\right)\xi + \frac{\bar{\gamma}-1}{2}(\pi^++\pi^-)- \frac{1}{\bar{M}}(\pi^+-\pi^-)=0,  
\end{align}%
which generalizes the compact-nozzle choking condition of \citet{Magri2016a} including changes in the specific-heat capacities. 
\section{Results}
The sound generated by the passage of a methane-air inhomogeneity through a supersonic nozzle with linear-velocity profile is investigated. 
First, the thermodynamic properties are evaluated from a one-dimensional counterflow-diffusion flame calculation, which is the basis of flamelet models in turbulent combustion \citep[e.g.,][]{WILLIAMS_BOOK1985}. 
This is to characterize the compositional inhomogeneity exiting the combustion chamber upstream of the nozzle \citep{Magri2016a}. 
Secondly,  the strengths of the acoustic sources $\bar{\Psi}$, $\bar{\aleph}$ \eqref{eq:dunnodd} and $\bar{\Phi}$ \eqref{eq:phistr} are evaluated. 
Thirdly, the transfer functions of the supersonic nozzle are calculated up to $He=0.5$ to quantify the indirect noise generated by the compositional blob.   

\subsection{Methane-air counterflow-diffusion flame}\label{sec:methane-air}
\begin{figure}
  \begin{center}
\includegraphics[width = 0.55\columnwidth,draft=false]{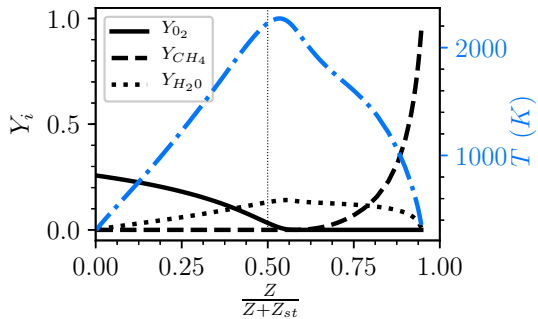}
   \caption{\label{fig:flamelet}One-dimensional pure-methane-air counterflow diffusion flame. The mixture-fraction, $Z$, is normalized such that the stoichiometric condition, $st$, occurs at 0.5 (thin dotted line). }
   \end{center}
\end{figure}

\begin{figure}
  \begin{center}
\includegraphics[width = 0.95\columnwidth,draft=false]{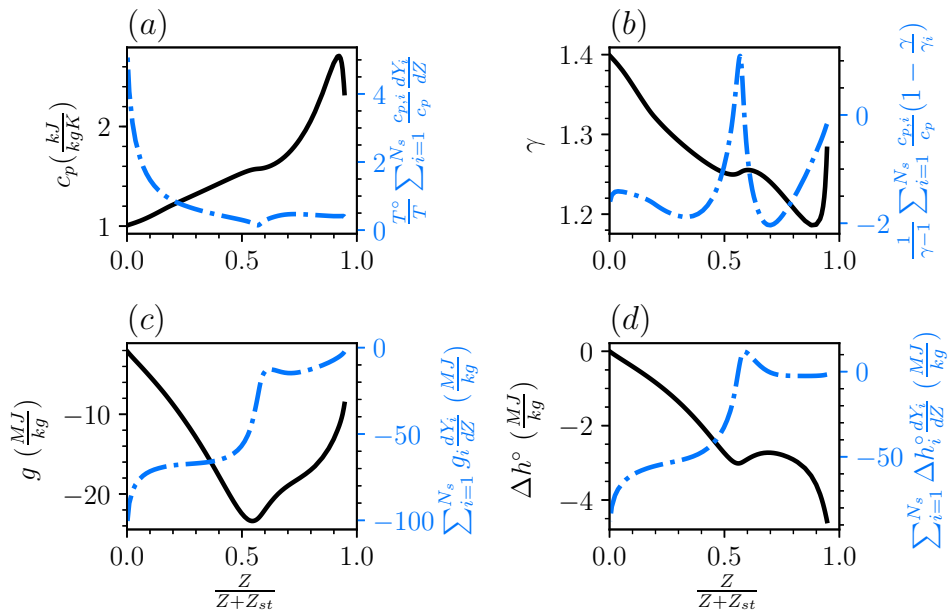}
   \caption{\label{fig:thermoprop}Thermodynamic properties necessary to characterize the acoustic sources $\bar{\Psi}$ and $\bar{\aleph}$ in \eqref{eq:dunnodd}, and $\bar{\Phi}$ in \eqref{eq:phistr}.}
   \end{center}
\end{figure}

The calculation is performed in the physical space with the Cantera library \citep{Cantera}.   
The reaction mechanism is GRI-Mech 3.0 \citep{GRI-30}, which consists of 53 species and 325 reactions and it is suitable for ideal-gas mixtures in natural-gas combustion. 
The fuel and oxidizer streams have temperature of $300$ (K) and pressure of $10^5$ (Pa).
The strain rate at stoichiometric condition is $18.7$ (1/s), which corresponds to a moderately-strained condition. 
The temperature and methane/oxygen/water mass fractions are shown in Figure~\ref{fig:flamelet}. 
The thermodynamic properties of the mixture, which are needed to evaluate the acoustic sources, $\bar{\Psi}$, $\bar{\aleph}$ and $\bar{\Phi}$, are shown in Figure~\ref{fig:thermoprop}. 
As for the heat-capacity factor, $\bar{\aleph}$, Figures~\ref{fig:thermoprop}(a,b) show that the contributions of variations in $c_p$ and $\gamma$ are comparable. 
As for the chemical potential function, $\bar{\Psi}$, Figure~\ref{fig:thermoprop}(c) shows that the largest contribution comes from the Gibbs' energy term, $\sum_{i=1}^{N_s}(\mu_i/W_i)dY_i/dZ = \sum_{i=1}^{N_s}g_idY_i/dZ$. 
The contribution from the formation-enthalpy term, $\sum_{i=1}^{N_s}\Delta h^\circ_idY_i/dZ$, is smaller, yet comparable (Figure~\ref{fig:thermoprop}(d)). 

At the mixture condition examined, $\bar{\Psi}$ and $\bar{\aleph}$ depend weakly on the strain rate.  
They vary by $\sim 8\%$ and $\sim1\%$, respectively, from near-equilibrium to near-quenching conditions (result not shown). 

\subsubsection{A note of caution} 
In the counter-flow diffusion flame calculation, $c_{p,i}$ is a function of the temperature, which, in turn, is a function of the mixture fraction, i.e., $c_{p,i}=c_{p,i}(T(Z))$. 
The multicomponent acoustic model for the nozzle assumes, however, that $c_{p,i}$ is constant (\S~\ref{sec:mathmodel}). 
%
%
%
Therefore, for the correct calculation of $\aleph$ and $\Phi$:
$d\log(c_p)/dZ$ in \eqref{eq:dunnodd} corresponds to $ \partial \log(c_p)/\partial Z = \sum_{i=1}^{N_s} (c_{p,i}/c_p)dY_i/dZ$ in the flamelet calculation, and 
$d\log(\gamma)/dZ$ in \eqref{eq:dunnodd},\eqref{eq:phistr} corresponds to  $\partial\log(\gamma)/\partial Z = \sum_{i=1}^{N_s} (c_{p,i}/c_p)\left(1- \gamma/\gamma_i\right)dY_i/dZ$.  
These relations ensure consistency with the calorically-perfect acoustic model. 
 
%
\subsection{Strength of acoustic sources and transfer functions}\label{sec:acsources}
\begin{figure}
  \begin{center}
\includegraphics[width = 0.95\columnwidth,draft=false]{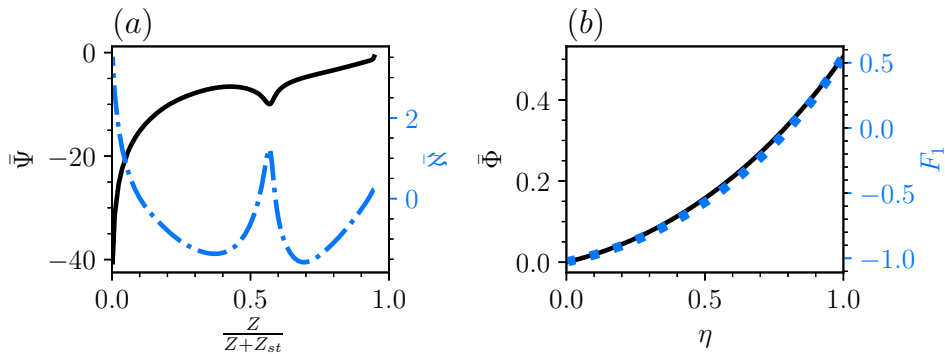}
   \caption{\label{fig:PsiFigure} Strength of the acoustic source, (a, left axis) $\bar{\Psi}$; (b, right axis), $\bar{\aleph}$; (b, left axis) $\bar{\Phi}$. The first component of $\mathbf{F}$ in \eqref{eq:fvect} is shown on the right axis of (b). The flow becomes sonic at $\eta=0.64$. }
   \end{center}
\end{figure}

The strengths of the acoustic sources, $\bar{\Psi}$ and $\bar{\aleph}$, are depicted in Figure~\ref{fig:PsiFigure}(a) as functions of the mixture fraction. 
At lean conditions close to $Z=0$, the chemical potential function, $\bar{\Psi}$, is the dominant source of indirect noise.
For other conditions, however, the heat-capacity factor, $\bar{\aleph}$, has the same order of magnitude as $\bar{\Psi}$. 

The condition considered at the nozzle's inlet is that of a mean-flow composition of $\bar{Z}=0.02$, which corresponds to a lean equivalence ratio of $0.358$ and $\bar{Z}/(\bar{Z}+Z_{st})=0.27$.
At this condition, $\bar{T}=1306.4$ (K)$, \bar{\gamma}=1.3$ and $\bar{c}_p=1288.8$ (J/(kg K)).  
The nozzle is supersonic and shock-free, with linear-velocity profile \citep{MARBLE_CANDEL_JSV1977}.
The Mach numbers are $\bar{M}_a=0.29$, at the inlet, and $\bar{M}_b=1.5$, at the outlet. 
The flow becomes sonic at $\eta=0.64$.
The stagnation pressure is $\bar{p}_0=1.06\times 10^5$ (Pa). 
The strengths of the acoustic sources are $\bar{\Psi}=-8.4$ and $\bar{\aleph}=-1.1$. 
The ratio at the inlet between the strength of entropy and compositional fluctuations in \eqref{eq:eq12} is $\bar{c}^{-1}_p/(\bar{\Psi}+\bar{\aleph}+\bar{\Phi}_a)=-8.1\times 10^{-5}$ (kg K/J). 
Interestingly, $\bar{\Psi}+\bar{\aleph}$ has opposite sign of an entropy fluctuation, $s'/\bar{c}_p$, generated by a hot spot of fluid. 

Figure~\ref{fig:PsiFigure}(b) shows the strength of the acoustic source, $\bar{\Phi}$, against the nozzle coordinate, and the first component of $\mathbf{F}$ in \eqref{eq:fvect}, which has been neglected in \S\ref{sec:solution}. 
Such a term can be ignored because its contribution averages out to a small number across the nozzle. 
The second component of $\mathbf{F}$ is $\sim O(10^{-2})$, hence, it is neglected. 

\begin{figure}
  \begin{center}
\includegraphics[width = 0.95\columnwidth,draft=false]{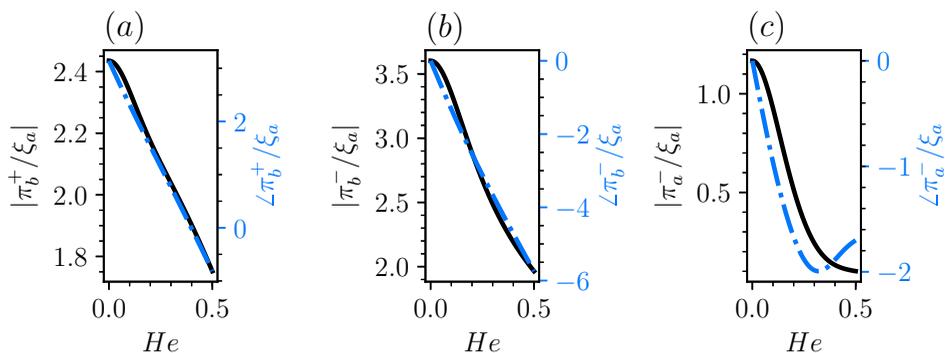}
   \caption{\label{fig:TFFigure} Transfer functions of compositional perturbations in a supersonic nozzle with linear-velocity profile. $\bar{M}_a=0.29$ and $\bar{M}_b=1.5$. Gain/phase on the left/right axes. }
   \end{center}
\end{figure}

In Figure~\ref{fig:TFFigure}, the transfer functions of compositional perturbations in a supersonic nozzle are shown. 
These are obtained with a second-order expansion of the propagator, which is performed at each point of a uniform spatial grid. 
The asymptotic solution matches the finite-difference solution of the boundary-value problem to a relative tolerance $\sim O(10^{-5})$ (result not shown). 
Gain/phase are depicted on the left/right axes. 
Panels (a,b) show the outgoing acoustics waves, $\pi^\pm_b$, which contribute to noise pollution. 
The finite spatial extent of the nozzle appreciably affects the transfer functions, whose gain monotonically and nonlinearly decreases with the Helmholtz number. 
The order of magnitude of the gain, however, remains unaltered.  
The phase has an almost-linear behaviour, as expected in a linear-velocity nozzle. 
Panel (c) shows the upstream-travelling acoustic wave, $\pi^-_a$, which is a quantity useful for thermo-acoustic stability analysis. 
Physically, the nozzle behaves similarly to a low-pass filter: higher Helmholtz numbers damp out the upstream-travelling acoustic wave. 
%
%
%
\subsection{Discussion}
First, the theoretical analysis, mathematical solutions and results presented may be useful to guide more experimental investigation and high-fidelity simulations to quantify compositional indirect noise. 
Secondly, the turbulent dissipation and dispersion of compositional blobs versus temperature inhomogeneities need to be quantified, for example, by estimating the effective diffusivity coefficient \citep[e.g.,][]{Wassmer2017}. 
Thirdly, an aspect to consider in experiments and real gas turbines is that compositional blobs cannot permeate non-porous walls, whereas temperature inhomogeneities can be affected by non-adiabatic boundaries. 
%
%
%
\section{Conclusions}
By observing that the gas exiting a combustion chamber of an aero-engine does not have homogeneous composition and specific heat capacities, an unsteady multicomponent gas acoustic model is proposed. 
It is shown that compositional inhomogeneities, named also ``compositional blobs", may generate noise through entropic  and isentropic excess-density mechanisms. 
By showing that the acoustic, entropic and compositional linear modes evolve independently in homogeneous media,  the hyperbolic partial differential equations are cast in invariant form. 
The multicomponent acoustic problem is mathematically solved by an asymptotic series with an infinite radius of convergence.
The proposed acoustic model and solution are able to predict the indirect noise generated by entropy spots and compositional blobs accelerated through subsonic and supersonic nozzles with a $4\times4$ (or $2\times2$) physics-based low-order model. 
This is applied to an inhomogeneous methane-air mixture entering a supersonic nozzle with linear-velocity profile. 
It is found that 
(i) at lean conditions, the chemical-potential function is the dominant acoustic source, of order $\sim O(10)$;
(ii) at lean-to-rich conditions, the chemical-potential function and the heat-capacity factor have the same order of magnitude, being $\sim O(1)$;
(iii) the compositional noise emitted by the nozzle contributes to indirect noise, whose order of magnitude is not sensitive to the Helmholtz number; and  
(iv) the acceleration of a compositional blob generates an upstream-travelling wave, which is sensitive to the Helmholtz number and appreciable in compact nozzles. 

The proposed analysis opens up new possibilities for separating and quantifying the sources of indirect combustion noise in aeronautical gas turbines. 
\section{Acknowledgements}
The author is supported by the Royal Academy of Engineering Research Fellowships Scheme. 
The author is grateful to Dr A. Giusti for his keen comments on the paper and A. K. Doan for helping set up the thermo-chemistry calculations. Discussions with Dr J. O'Brien, Dr A. Agarwal and Prof M. Ihme are gratefully acknowledged. The author thanks Prof S. Hochgreb and E. Rolland for discussions on their experimental data. 
\bibliographystyle{jfm}
\end{document}